\documentclass[usenatbib]{svjour2}

\usepackage{amsmath, amssymb, amsfonts}
\usepackage{txfonts}
\usepackage{graphicx}

\usepackage{array}
\def\etal{{\it et~al. }} 
\usepackage[dvips]{color}
\usepackage{ulem}
\usepackage{natbib}

\begin{document}

\title{Dynamics of planets in retrograde mean motion resonance} 

\author{Julie Gayon, Eric Bois, \& Hans Scholl}
\institute{Nice Sophia-Antipolis University, CNRS, Observatoire de la C\^ote
d'Azur, Laboratoire\linebreak Cassiop\'ee, B.P. 4229, F-06304 Nice Cedex 4,
France\\ 
\email{Julie.Gayon@oca.eu}}

\date{}

\maketitle

\begin{abstract}\hspace{2mm}

In a previous paper (Gayon \& Bois 2008a), we have shown the general 
efficiency of retrograde resonances for stabilizing compact planetary
systems. Such retrograde resonances can be found when two-planets of a 
three-body planetary system are both in mean motion resonance and revolve in 
opposite directions. For a particular two-planet system, we have
also obtained a new orbital fit involving such a counter-revolving 
configuration and consistent with the observational data.

In the present paper, we analytically investigate the three-body problem in
this particular case of retrograde resonances. We therefore define a new set 
of canonical variables allowing to express correctly the resonance angles and 
obtain the Hamiltonian of a system harboring planets revolving in opposite 
directions.
The acquiring of an analytical ``rail'' may notably contribute to a deeper
understanding of our numerical investigation and provides the major
structures related to the stability properties. A comparison between our 
analytical and numerical results is also carried out.

\end{abstract}

\newpage
\section{Introduction}
The simplicity and elusiveness of the full three-body problem have captived
mathematicians for centuries. Still today, because of its
non-integrability, the three-body problem remains tricky. 
It has been solved for particular cases, by reducing the number of
degrees of freedom. One of the first major results is the determination of 
families of periodic orbits in the three-body problem under certain
assumptions (see for instance H\'enon 1976 or Hadjidemetriou 1976).

Since the first observations of extrasolar systems, the interest of the
general three-body problem has been boosted, mainly under the
coplanar assumption. Moreover, the occurrence of bodies in Mean 
Motion Resonances (MMR) allowed to study the stability of several
two-planet systems (see e.g. Hadjidemetriou 2002, 2008 or Callegari 
\etal 2004). The secular behavior of two-planet
systems has been investigated as well (e.g. Henrard \& Libert 2008). 
Because most presently known extrasolar planets move on eccentric
orbits, Beaug\'e \& Michtchenko (2003) investigated the eccentric, coplanar
three-body problem to characterize the dynamics of such two-planet systems.

Faced to the detection of multi-planetary systems (one star and 2 planets or
more), and to the increasing number of degrees of freedom, numerical methods
of global analysis have been performed in order to study their
  dynamical stability. By combining {\it N}-body
integrators with numerical tools for detection of chaos (notably
fast indicators or frequency analyses), the exploration of the {\it N}-body 
planetary problem (with {\it N} $\ge 3$) is accessible from numerical
methods. As a consequence, mechanisms generating stability of multi-planetary
systems can be identified and explained. For instance, mechanisms 
involving a MMR as well as an Apsidal Synchronous
Precession\footnote{In case of ASP, the apsidal lines on average precess at
  the same rate.} (ASP hereafter) have been intensively studied  (e.g. Lee \&
Peale 2002, 2003; Bois \etal 2003, Ji \etal 2003). A solution involving both
MMR and ASP describes well the stability of eccentric, compact
multi-planetary systems, but may not however be unique. We note, for example, 
that other multi-planetary systems have been found to be mainly controlled by 
secular dynamics (e.g. Michtchenko \etal 2006 or Ji \etal 2007).
In Gayon \& Bois (2008a), by using a numerical method of global
analysis,\footnote{We used the MIPS (Megno Indicator for Planetary Systems)
method based on the MEGNO technique (Mean Exponential Growth factor of Nearby
Orbits; see Cincotta \& Sim\`o 2000).} we found a novel stabilizing mechanism 
involving a {\it retrograde} MMR in a two-planet system, which means that the
two planets are both in MMR and {\it revolve in opposite 
direction}.\footnote{For instance, the inner planet has a prograde direction
while the outer planet a retrograde one. In the following, we
  will also use the terminology of ``counter-revolving configuration''.} 
Let us note that the difference between a {\it prograde} MMR
and a {\it retrograde} MMR does not only lie in the retrograde motion
of one of the two planets. The mechanisms of stability related
to a retrograde MMR and their underlying properties differ from the
prograde case. Such stabilities are generally more robust (see Gayon \& Bois
2008a, 2008b).

The assumption that two giant planets are in a MMR and revolving in opposite 
directions around their hosting 
star is apparently contradicting to the most accepted formation theory of 
planetary systems, notably to the formation and evolution of the 
resonant planetary systems (core accretion mechanism combined by a planetary 
migration scenario). However, in Gayon \& Bois (2008a), we present two 
feasible processes leading to planets revolving in opposite directions. 
The first scenario 
has been introduced by Nagasawa \etal (2008) and consists in the combination 
of a planet-planet scattering, a tidal circularization and Kozai 
mechanisms. Starting from a hierarchical 3-planet system and considering a 
migration mechanism including a process of planet-planet scattering and tidal 
circularization, the authors show that close-in planets may be  formed. In a 
few cases, due to the Kozai mechanim, one planet enters a retrograde motion. 
The second feasible process imagined by Varvoglis (private discussions) and 
ourselves is related to the capture of free-floating planets. By 
integrating the trajectories of planet-sized bodies that encounter a coplanar 
two-body system (a Sun-like star and a Jupiter mass), Varvoglis
finds that the probability of capture is significant, and,
morever, that the percentage of free-floating planets captured is higher for
retrograde motions than for prograde motions. As a consequence, it would be
possible to find one day some planetary systems in  counter-revolving
configuration. 

In the case of a two-planet system, analytical and numerical methods are
complementary. Besides, the acquiring of an analytical ``rail'' contributes to 
a deeper understanding of the numerical exploration. As a result, the
analytical approach permits to study thoroughly the numerical
results. By plotting surfaces of sections, it
provides the major structures related to the stability properties. 
As a consequence, in the present paper, we propose a Hamiltonian
expansion based on this novel and feasible stabilizing mechanism
introduced in Gayon \& Bois (2008a and 2008b).
We therefore investigate the three-body problem in the particular
case of retrograde resonances, adapting the Hamiltonian approach of Beaug\'e
\& Michtchenko (2003) firstly expanded for eccentric, coplanar, and {\it
prograde} orbits. From this analytical expansion, we can correctly 
derive the expression of resonance angles in the case of a retrograde MMR. 
A comparison with the numerical results obtained in Gayon \& Bois (2008a) is
also carried out.

In the present paper, we firstly introduce the general expressions of the
Hamiltonian as well as the canonical variables in the case of two massive
planets moving around their host star (Section 2). We note that a dynamical 
system composed of two satellites orbiting a planet is also an equivalent
problem. All the results shown in the present paper may therefore be applied 
to satellite systems. The consideration of planets revolving in opposite
directions involves
new expressions for canonical variables presented in Section 3. We adapt the
disturbing function of the Beaug\'e \& Michtchenko model to retrograde motions
in Section 4. In Section 5, we express the Hamiltonian for retrograde
resonances. An orbital fit of the HD\thinspace73526 planetary
system involving a 2:1 retrograde MMR was found consistent with the
observational data in Gayon \& Bois (2008a). As a consequence, in Section 6, 
we apply the analytical model to the HD\thinspace73526 planetary system
ruled in such conditions and compare our numerical and 
analytical results. Finally, in Section \ref{sec_surfsec}, our results 
obtained by the analytical expansion are visualized by surfaces of section.

\newpage
\section{General expression of the Hamiltonian}
We suppose a star of mass $M_0$ and 2 planets of mass $m_1$ and $m_2$ 
such as\linebreak $M_0 \gg m_1, m_2$ and moving around their barycenter. 
We choose to express the Hamiltonian of the system in the heliocentric frame
by using the following Delaunay variables~: 

\begin{equation} \label{var_Delaunay}
  \begin{aligned}
     M_i \hspace{1cm}& L_i& = & \hspace{0.2cm} \beta_i \sqrt{\mu_i a_i} \\
     {\omega}_i \hspace{1cm}& G_i & = & \hspace{0.2cm} L_i \sqrt{1 - e_i^2}\\
     \Omega_i \hspace{1cm}& H_i & = & \hspace{0.2cm} G_i \cos I_i
  \end{aligned}
\end{equation}
with $a_i$,$e_i$,$I_i$,$\Omega_i$,$\omega_i$, and $M_i$ the geometrical 
orbital elements of the planet $i$, and $\mu_i = G \sqrt{M_0+m_i}$. G is the
gravitational constant. $\beta_i=M_0 \, m_i / (M_0+m_i)$ is the
reduced mass for the planet $i$. 

The Hamiltonian of the system (denoted $F$) consists of a
Keplerian part ($F_0$) and a disturbing function ($F_1$)~: 
\begin{equation}
F = F_0 + F_1 \hspace{0.5cm} \textrm{with} \hspace{0.5cm} \left\{   
\begin{aligned}
F_0 &= - \sum_{i=1}^{2} \frac{\mu_i^2\beta_i^3}{2 L_i^2}\\
F_1 &= -G m_1 m_2 \frac{1}{\Delta} + \frac{m_1 m_2}{M_0} (\dot{x}_1\dot{x}_2 
                                   + \dot{y}_1\dot{y}_2 + \dot{z}_1\dot{z}_2) 
    \end{aligned} \right.
\label{eq_Ham}
\end{equation}
The leading part of the disturbing function, called the direct part, depends 
on the instantaneous distance between the two planets ($\Delta$). Owing to the
choice of the origin of the coordinate systems, the second part of $F_1$ is
called the indirect part of the disturbing function. $F_1$ depends on the 
barycentric velocities $(\dot{x}_i, \dot{y}_i, \dot{z}_i)$ of each planet,
expressed in cartesian coordinates.

To study the 3-body problem, different sets of canonical variables can
be used, depending on the simplifications carried out for reducing the number 
of degrees of freedom. For instance, Beaug\'e \& Michtchenko (2003) applied 
the so-called modified Delaunay variables to solve the coplanar, prograde and
eccentric 3-body problem. In the present paper, we define a new set of
canonical variables, especially devoted to planets revolving in opposite
directions.

\section{Set of canonical variables for planets revolving in opposite directions}
In this study, we choose to set a prograde direction of revolution for the 
inner planet (noted 1) and a retrograde one for the outer planet (noted 2). 
From the previous set of Delaunay variables (\ref{var_Delaunay}), we 
therefore define the two following sets of canonical variables for planets 1 
and 2 respectively~: 
\begin{equation} \label{var_Poincare}
  \left\{
\begin{aligned}
     \lambda_1&=M_1+\tilde{\omega}_1 \hspace{1cm}& L_1&& = & \hspace{0.2cm} \beta_1 \sqrt{\mu_1 a_1} \\
     -\tilde{\omega}_1&=-(\Omega_1+\omega_1) \hspace{1cm}& L_1& - G_1 & = &
     \hspace{0.2cm} L_1 (1-\sqrt{1 - e_1^2})\\
     -\Omega_1& \hspace{1cm}& G_1& - H_1 & = & \hspace{0.2cm} L_1 \sqrt{1 -
       e_1^2} (1- \cos I_1)
\end{aligned}
  \right.
\end{equation}
\vspace{5mm}
\begin{equation} \label{var_Poincare_retro}
   \left\{
  \begin{aligned}
     \lambda_2&=-M_2+\tilde{\omega}_2 \hspace{1cm}& -L_2&& = & - \hspace{0.1cm} \beta_2 \sqrt{\mu_2 a_2} \\
     -\tilde{\omega}_2&=-(\Omega_2-\omega_2) \hspace{1cm}& G_2& - L_2 & = &
      -L_2 (1 - \sqrt{1 - e_2^2})\\
     \Omega_2& \hspace{1cm}& G_2& + H_2 & = & \hspace{0.4cm} L_2 \sqrt{1 - e_2^2} (1 + \cos I_2)
\end{aligned}
\right.
\end{equation}%\\

The first set of canonical variables (\ref{var_Poincare}) corresponds to 
the well-known modified Delaunay canonical variables while the second one
(\ref{var_Poincare_retro}) is developed for retrograde planetary motion.
$\lambda_i$ is the mean longitude of the planet $i$.
We note that the longitude of periastron is defined by: 
$\tilde{\omega} = \Omega + \omega$, 
for a coplanar and {\it prograde} planetary motion 
(which is equivalent to consider $I=0^\circ$) and by
$\tilde{\omega} = \Omega - \omega$, for a coplanar and {\it retrograde} 
planetary motion (equivalent to $I=180^\circ$). Restricting the study to a 
coplanar but counter-revolving problem, the set of canonical variables of the
system can be simplified as follows~:
\begin{equation}
  \begin{aligned}
     \lambda_1         \hspace{1cm}& L_1 \\
     \lambda_2         \hspace{0.8cm}-& L_2\\
     -\tilde{\omega}_1 \hspace{1cm}& L_1 - G_1\\ 
     -\tilde{\omega}_2 \hspace{1cm}& G_2 - L_2 
  \end{aligned}
\end{equation}

%\newpage
\section{Expansion of the disturbing function}
The method used for the expansion of the disturbing function originates from
Beaug\'e \& Michtchenko (2003). They expanded an expression for the
Hamiltonian that is valid for high eccentricities of one or both
planets. Their method provides a good assessment of the disturbing
function when planets are close to the collision point. Hence, the solution 
converges in all points of the phase space except for the singularities
(corresponding to planetary collisions). Moreover, the rate of convergence
does not depend on the values of eccentricities but rather on the
order of magnitude of the
distribution function itself. Besides, depending on the studied planetary
system, even for very eccentric orbits, the rate of convergence of the
disturbing function may be relatively fast. 

Classical methods such as Laplace 1799 or Kaula 1962 were expanded for coplanar
and quasi-circular orbits and were useful for studying dynamics of the Solar
System (or hierarchical systems in general). Nonetheless, due to the detection
of eccentric exo-planetary systems, new and specific expansions of the 
disturbing function were needed. As a consequence, the Beaug\'e \& Michtchenko 
method seems much more suitable for such eccentric planetary orbits.
To apply this method for counter-revolving configurations, we propose 
to expand again the disturbing function, by considering the prograde 
motion of the inner planet and the retrograde motion of the outer planet.

\subsection{Direct part of the disturbing function}
The direct part of the disturbing function that generally encounters problems
of convergence may be expressed as a function of the heliocentric radial
distances $r_i$ of both planets and the $\psi$ angle between both bodies as seen
from the star~:
\begin{equation}
\frac{1}{\Delta} = (r_1^2 + r_2^2 - 2 r_1 r_2 \cos \psi)^{-1/2}
\end{equation}
We note that for counter-revolving configurations, the angle $\psi$ between
both bodies is defined by\footnote{For both prograde orbits, the $\psi$ angle
is defined by: $\psi=f_1-f_2+\Delta\tilde{\omega}$ with
$\tilde{\omega}_i=\Omega_i+\omega_i$.}~:  
\begin{equation}
\psi = f_1 + f_2 + \Delta{\tilde{\omega}}  
\label{eq_S}
\end{equation}
with $f_i$ the true anomaly of the planet $i$,
$\Delta\tilde{\omega}=\tilde{\omega}_1 - \tilde{\omega}_2$ and
$\tilde{\omega}_{\{1,2\}}$ defined in (\ref{var_Poincare}) and
(\ref{var_Poincare_retro}). 
A concise expression of the previous equation is written in 
(\ref{eq_rho}) by taking into account the ratio $\rho=r_1/r_2$ such as~:
\begin{equation}
\frac{r_2}{\Delta} = (1 + \rho^2 - 2 \rho \cos \psi)^{-1/2}
\label{eq_rho}
\end{equation}
The key of the Beaug\'e \& Michtchenko (2003) method lies in the expansion in 
power series in a new variable noted $x$ and corresponding to a measurement of 
the proximity of the initial condition to the singularity in $1/\Delta$~:
\begin{equation}
\frac{r_2}{\Delta} = (1+x)^{-1/2} \simeq \sum_{n=0}^N b_n x^n
\label{eq_R2D}
\end{equation}
with $x=\rho^2 - 2 \rho \cos \psi$. $r_2/\Delta$ has a singularity at
$x=-1$. The determination of the coefficients $b_n$ is performed by the way of 
a linear regression for $x$ values greater than $-1+\delta$, $\delta$ being a
positive parameter close to zero. A good precision of the direct part of the
disturbing function may be reached for a good compromise between the $\delta$
value and the choice of $N$ order in the series expansion. Contrary to 
classical methods involving Fourier series of the $\psi$ variable or power
series in $\rho$, not only the convergence rate of this method is improved but
also the expansion of the disturbing function can be applied for eccentric
two-planet systems. More details can be found in Beaug\'e \& Michtchenko
(2003). 

From (\ref{eq_R2D}) and by using the explicit expression of $x$, we
find~: 
\begin{equation}
\frac{r_2}{\Delta} \simeq \sum_{l=0}^{N} \sum_{k=0}^{l} \, b_l \, (-2)^k \,
  {l \choose k} \, \rho^{2l-k} \, \cos^k \psi
\end{equation}
Changing from powers of $\cos \psi$ to multiples of $\psi$ and by using the
explicit expression of the $\psi$ angle, one obtains~:
\begin{equation}
\begin{aligned}
\frac{a_2}{\Delta} & \simeq \sum_{l=0}^{N} \sum_{u=0}^{N-l} 2\, A_{l,u}\, 
\alpha^{2u+l} \left( \frac{r_1}{a_1} \right)^{2u+l} \left( \frac{r_2}{a_2}
\right)^{-2u-l-1} \cos \, (l f_1+ l f_2+ l\Delta\tilde{\omega})
\end{aligned}  
\end{equation}
with $A_{l,u} = (-1)^l \sum_{t=u}^{\min(2u,N-l)}\, b_{l+t}\, {l+t \choose
  l+2t-2u}\, {l+2t-2u \choose t-u}\, \gamma_l$, $\gamma_l =$ 
$\left\{ \begin{aligned} 1/2 \textrm{ if } l=0\\ 
 1 \textrm{ if } l > 0 \end{aligned} \right.$

The direct part of the disturbing function may be expressed in terms of mean 
anomaly by using the Fourier expansion of the following functions (e.g. Hughes
1981)~:  
\begin{equation}
   \begin{aligned}
\left( \frac{r}{a} \right) ^n &\cos(l f) &= \sum_{m=-\infty}^{\infty}
  X_m^{n,l} \cos(m M)\\ 
\left(\frac{r}{a}\right)^n &\sin(l f) &= \sum_{m=-\infty}^{\infty} 
  X_m^{n,l} \sin(m M) 
   \end{aligned}
\label{eq_hansen}
\end{equation}
with $X_m^{n,l}$ the Hansen coefficient function of the
eccentricity (Kaula 1962)~:
\begin{equation}
X_m^{n,l} = e^{|l-m|} \sum_{s=0}^{\infty} Y_{s+w_1 , s+w_2}^{n,l} e^{2s}
\label{eq_newcomb}
\end{equation}
with $Y_{s+w1,s+w2}^{n,l}$ the Newcomb operators, $w_1=\max(0,m-l)$ and
$w_2=\max(0,l-m)$.\linebreak Let us recall that the Newcomb operators obey to
simple recurrence relations (Brouwer \& Clemence 1961; Murray \& Dermott
1999).

Substituting (\ref{eq_newcomb}) into (\ref{eq_hansen}), we obtain~: 
\begin{equation}
   \begin{aligned}
\left(\frac{r}{a}\right)^n &\cos(lf) &= \sum_{j=0}^{\infty}
\sum_{m=-\infty}^{\infty} B_{n,l,j,m} \, e^j \cos(mM)\\
\left(\frac{r}{a}\right)^n &\sin(lf) &= \sum_{j=0}^{\infty}
\sum_{m=-\infty}^{\infty} B_{n,l,j,m} \, e^j \sin(mM)
   \end{aligned}
\label{eq_ra}
\end{equation}

The direct part of the disturbing function is therefore expressed as follows~:
\begin{equation}
   \begin{aligned}
\frac{a_2}{\Delta} & \simeq \sum_{j,k=0}^{\infty}
\sum_{m,n=-\infty}^{\infty} \sum_{l=0}^{N} \sum_{i=0}^{2N} A_{l,(i-l)/2} \, 
D_{i,l,j,k,m,n} \, \alpha^{i} \, e_1^j \, e_2^k \cos(m M_1 + n M_2 + l
\Delta\tilde{\omega})
   \end{aligned}
\label{eq_direct}
\end{equation}
with $D_{i,l,j,k,m,n} = 2 \, B_{i,l,j,m} \, B_{-i-1,l,k,n}$  and 
$B_{i,l,j,m} = Y_{\frac{j-|l-m|}{2}+w_1,\frac{j-|l-m|}{2}+w_2}^{i, l}$.

\subsection{Indirect part of the disturbing function}
From (\ref{eq_Ham}), the indirect part of the disturbing function is
given by the function $T_1$~:
\begin{equation}
T_1 = \frac{m_1 m_2}{M_0} (\dot{x}_1\dot{x}_2 + \dot{y}_1\dot{y}_2 +
\dot{z}_1\dot{z}_2) 
\end{equation}
where $\dot{x}_i$, $\dot{y}_i$, and $\dot{z}_i$ are the barycentric velocities
of the planet $i$, expressed in cartesian coordinates. 
In the case of coplanar orbits, the expressions of $\dot{x}_i$ and $\dot{y}_i$
take into account the direction of motion of the planet $i$. We
define $\dot{x}_i$ for a prograde motion and a retrograde one as
follows\footnote{A similar equation is obtained for the expression of
$\dot{y}_i$.}~: 
\begin{equation}
\dot{x}_i = \frac{dx_i}{dt} = \frac{\partial{x_i}}{\partial{M_i}}
\frac{d M_i}{dt} = 
\left\{ \begin{aligned}
 &\frac{\partial x_i}{\partial M_i} n_i&\textrm{for a progade motion}\hspace{2mm}\\
-&\frac{\partial x_i}{\partial M_i} n_i&\textrm{for a retrogade motion}
\end{aligned}\right.
\end{equation}

 \noindent Consequently, considering a coplanar problem and planets revolving
 in opposite directions, the indirect part is equal to~: 
\begin{equation}
   \begin{aligned}
T_1 = -\frac{G m_1 \, m_2}{a_2} \alpha^{-1/2} 
  \left[ \frac{\partial}{\partial M_1} \left( \frac{x_1}{a_1}  \right)
 \frac{\partial}{\partial M_2} \left( \frac{x_2}{a_2}  \right)
 + \frac{\partial}{\partial M_1} \left( \frac{y_1}{a_1}  \right)
 \frac{\partial}{\partial M_2} \left( \frac{y_2}{a_2}  \right) \right]
    \end{aligned}
\label{eq_T1}
\end{equation}

\noindent By using (\ref{eq_ra}) and considering the prograde motion of
planet 1 and the retrograde one of planet 2, one obtains~:
\begin{equation}
   \begin{aligned}
\frac{x_1}{a_1} & = \left( \frac{r_1}{a_1} \right) \cos(\tilde{\omega}_1 + f_1)
 = \sum_{j=0}^{\infty} \sum_{m=-\infty}^{\infty} B_{1,1,j,m}
e_1^j \cos(\tilde{\omega}_1 + m M_1)\\
\frac{y_1}{a_1} & = \left( \frac{r_1}{a_1} \right) \sin(\tilde{\omega}_1 + f_1)
  = \sum_{j=0}^{\infty} \sum_{m=-\infty}^{\infty} B_{1,1,j,m}
e_1^j \sin(\tilde{\omega}_1 + m M_1)\\
\frac{x_2}{a_2} & = \left( \frac{r_2}{a_2} \right) \cos(\tilde{\omega}_2 - f_2)
 = \sum_{k=0}^{\infty} \sum_{n=-\infty}^{\infty} B_{1,1,k,n}
e_2^k \cos(\tilde{\omega}_2-n M_2)\\
\frac{y_2}{a_2} & = \left( \frac{r_2}{a_2} \right) \sin(\tilde{\omega}_2 - f_2)
  = \sum_{k=0}^{\infty} \sum_{n=-\infty}^{\infty} B_{1,1,k,n}
e_2^k \sin(\tilde{\omega}_2 - n M_2)
   \end{aligned}
\end{equation}

\noindent Substituting these results into (\ref{eq_T1}), we find~:
\begin{equation}
   \begin{aligned}
 T_1 = \frac{G m_1 m_2}{a_2} &\sum_{j,k=0}^{\infty}
\sum_{m,n=-\infty}^{\infty} \sum_{i=0}^{2N} \bar{A}_i \alpha^i\\
&\times m n B_{1,1,j,m} B_{1,1,k,n}
e_1^j e_2^k \cos(m M_1 + n M_2 + \Delta\tilde{\omega})
   \end{aligned}
\label{eq_indirect}
\end{equation}
where $\bar{A}_i$ are constant coefficients such as~:
$\alpha^{-1/2} = \sum_{i=0}^{2N} \bar{A}_i \alpha^i$.\\
\\

By combining (\ref{eq_direct}) and (\ref{eq_indirect}), we 
obtain the complete expression of the disturbing function as follows~:
\begin{equation}
   \begin{aligned}
F_1 = -\frac{G m_1 m_2}{a_2} \sum_{j,k=0}^{\infty} \sum_{m,n=-\infty}^{\infty}
\sum_{l=0}^{N} \sum_{i=0}^{2N} R_{i,j,k,m,n,l} \,
  \alpha^i \, e_1^j \, e_2^k \cos(m M_1 +n M_2 + l
 \Delta\tilde{\omega})
   \end{aligned} 
\end{equation}
where $R_{i,j,k,m,n,l} = A_{l,(i-l)/2} \, D_{i,,l,j,k,m,n} - \delta_{l,1} \,
\bar{A_i} m \,  n \, 
B_{1,1,j,m} \, B_{1,1,k,n}$ are constant coefficients. $R_{i,j,k,l,m,n,l}$ are
independent of initial conditions and then require to be determined once. We 
are now able to express the angles of resonance in the case of
counter-revolving configurations as shown in the following section.

\section{The resonant average Hamiltonian}
Considering planets close to a MMR and revolving in opposite directions, we 
define the MMR ratio by $p+q/p$ with $p \ne 0$ {\it and} $p < 0$. For
instance, when two planets revolve in opposite direction and have a period
ratio of $2$, the $q$ order of resonance is equal to $3$ and $p=-1$. We set
$s=p/q$ and define the following set of canonical variables in the case of a
retrograde MMR~: 
\begin{equation}\label{eq_average}
   \begin{aligned}
   \lambda_1 & 
   & J_1 =& \hspace{4mm}L_1 + s (I_1 + I_2)\\
   \lambda_2 & 
   & J_2 =& -L_2 - (1+s) (I_1 + I_2)\\
   \sigma_1  & = (1+s) \lambda_2 - s \lambda_1 - \tilde{\omega}_1 \hspace{1cm}
   & I_1 =& \hspace{4mm}L_1\, (1 - \sqrt{1-e_1^2})  \\
   \sigma_2 &= (1+s) \lambda_2 - s \lambda_1 - \tilde{\omega}_2  
   & I_2 =& -L_2\, (1-\sqrt{1-e_2^2}) 
   \end{aligned}
\end{equation}
where $\sigma_1$ and $\sigma_2$ are the resonant angles while $I_1$
and $I_2$ are the conjugate momenta depending on the eccentricities $e_1$ and
$e_2$.\\ 

Let $\theta$ be the following angle of the disturbing function such as~: 
$\theta = m M_1 + n M_2 + l \Delta\tilde{\omega}$. Considering the new set of
variables, the expression of $\theta$ becomes~:
\begin{equation}
   \begin{aligned}
\theta  = m \sigma_1 - n \sigma_2 + l (\sigma_2 - \sigma_1) + [m(p+q)-np]Q
   \end{aligned}
\end{equation}
where $qQ=(\lambda_1-\lambda_2)$ is the synodic angle. Hence, the disturbing
function depends on three angular variables $(\sigma_1, \sigma_2,
Q)$. Considering the new set of canonical variables $(\sigma_1, \sigma_2,
Q, \lambda_2 ; I_1, I_2, qJ_1, J_1+J_2)$, we find the following constant
of motion~:
\begin{equation}
J_1 + J_2 = \textrm{constant}
\end{equation}

As a consequence, our system has therefore two integrals of motion: the 
$F$ Hamiltonian and
$J_{tot}=J_1+J_2$. It is well known that the frequency of the angle $Q$ is
much higher than that of $\sigma_i$. As a consequence, we consider only
long period perturbations in the Hamiltonian. The system is then averaged with
respect to the synodic angle as follows~:
\begin{equation}\label{eq_Q}
\bar{F}_1 = \frac{1}{2\pi} \int_0^{2\pi} F_1 dQ
\end{equation}
The averaging over the $Q$ variable written in (\ref{eq_Q}) implies a third 
constant of motion, namely $J_1$. As a consequence, the system consists of 
four degrees of freedom and three constants of motion.
From now on, the system can be reduced to four independent variables, namely 
$\sigma_1, \sigma_2, I_1$ and $I_2$.
Taking into account the condition of MMR
($n=m(p+q)/p$), we obtain the final expression of the disturbing 
function (\ref{final_disturb}) and the average Hamiltonian
(\ref{final_Ham}) respectively, for a retrograde resonance~:
\begin{equation} \label{final_disturb}
   \begin{aligned}
\bar{F}_1  = -\frac{G m_1 m_2}{a_2} &\sum_{j=0}^{j_{max}}
\sum_{k=0}^{k_{max}} \sum_{m=-m_{max}}^{m_{max}} \sum_{l=0}^{l_{max}} \sum_{i=0}^{2N}
\bar{R}_{i,j,k,m,l} \\
& \times \alpha^i e_1^j e_2^k \cos((m-l)\sigma_1 + (l-n) \sigma_2)
   \end{aligned}
\end{equation}

\begin{equation} \label{final_Ham}
\bar{F} = -\sum_{i=1}^{2} \frac{\mu_i^2 \beta_i^3}{2 L_i^2} 
      - \frac{G m_1 m_2}{a_2} \sum_{i,j,k,m,l} \bar{R}_{i,j,k,m,l} \, \alpha_i \,
      e_1^j \, e_2^k \, \cos((m-l)\sigma_1 + (l-n) \sigma_2)
\end{equation}

Due to D'Alembert's properties of the disturbing function, some 
coefficients are null if one (or more) of the following conditions is
reached~:  1) $j<|m-l|$, 2) $k<|l-n|$, 3) $(m-l)$ even (odd) number and j odd
(even) number, 4) $(l-n)$ even (odd) number and k odd (even) number.

\section{Comparison with numerical methods}\label{compare}
Using the final expression of the Hamiltonian, the $J_1$ and $J_2$ constants
of motion as well as the following set of canonical variables, 
\begin{equation}
\begin{aligned}
\sigma_1 & = (1+s) \lambda_2 - s \lambda_1 - \tilde{\omega}_1 
         & \hspace{2mm}  \hspace{5mm} &
     I_1 = \hspace{2mm}L_1 (1 - \sqrt{1-e_1^2})\\
\sigma_2 & = (1+s) \lambda_2 - s \lambda_1 - \tilde{\omega}_2 
         & \hspace{2mm}  \hspace{5mm} &
     I_2 = -L_2 (1 - \sqrt{1-e_2^2})
\end{aligned}
\end{equation}
we can firstly integrate a two-planet system both in counter-revolving
configuration and in MMR and secondly, compare the analytical results with
numerical ones. Using a Bulirsch-Stoer method, we integrate 
numerically the following differential equations~: 
\begin{equation}\label{eq_derivatives}
\begin{aligned}
\dot{I_i} &=& - \frac{\partial \bar{F}}{\partial \sigma_i} &= &-\frac{\partial
  \bar{F}_1}{\partial \sigma_i}&\Huge{\textrm{}}\\
%\textrm{\tiny{.}}
\dot{\sigma_i} &=& \frac{\partial \bar{F}}{\partial I_i} &= &
\frac{\partial \bar{F}_1}{\partial I_i}&
+\frac{\partial F_0}{\partial I_i}
\end{aligned}
\end{equation}

For the integration of (\ref{eq_derivatives}), we use our initial conditions 
for the HD$\thinspace$73526 system, located very close to the 2:1 retrograde
MMR (i.e. at the edge of a V-shape structure in a stability map in $[a,e]$
orbital elements; see Gayon \& Bois, 2008a)~:
$$M_0 =1.08 \, M_\odot $$
\begin{equation}\label{eq_CI}
\begin{aligned}
M_1 & = 2.9 \, M_{Jup} & \hspace{2cm}  M_2 & = 2.5 \, M_{Jup}\\
a_1 & = 0.66 \textrm{ AU} & \hspace{2cm}  a_2 & = 1.05 \textrm{ AU}\\
e_1 & = 0.19 & \hspace{2cm}  e_2 & = 0.14\\
\sigma_1 & = 94 \textrm{ (deg)} & \hspace{2cm}  \sigma_2 & = 94 \textrm{ (deg)}
\end{aligned}
\end{equation}
Such initial conditions 
(\ref{eq_CI}) are sufficiently close to the retrograde MMR to apply our 
analytical expansion.\footnote{We forecast to study various initial conditions
in a forthcoming paper, notably for a ``full'' retrograde resonance.}
These initial conditions (more paticularly the semi-major axes) are averaged 
for the analytical expansion.

\begin{figure}[!h]
    \centering
     \includegraphics[angle=270,scale=0.28]{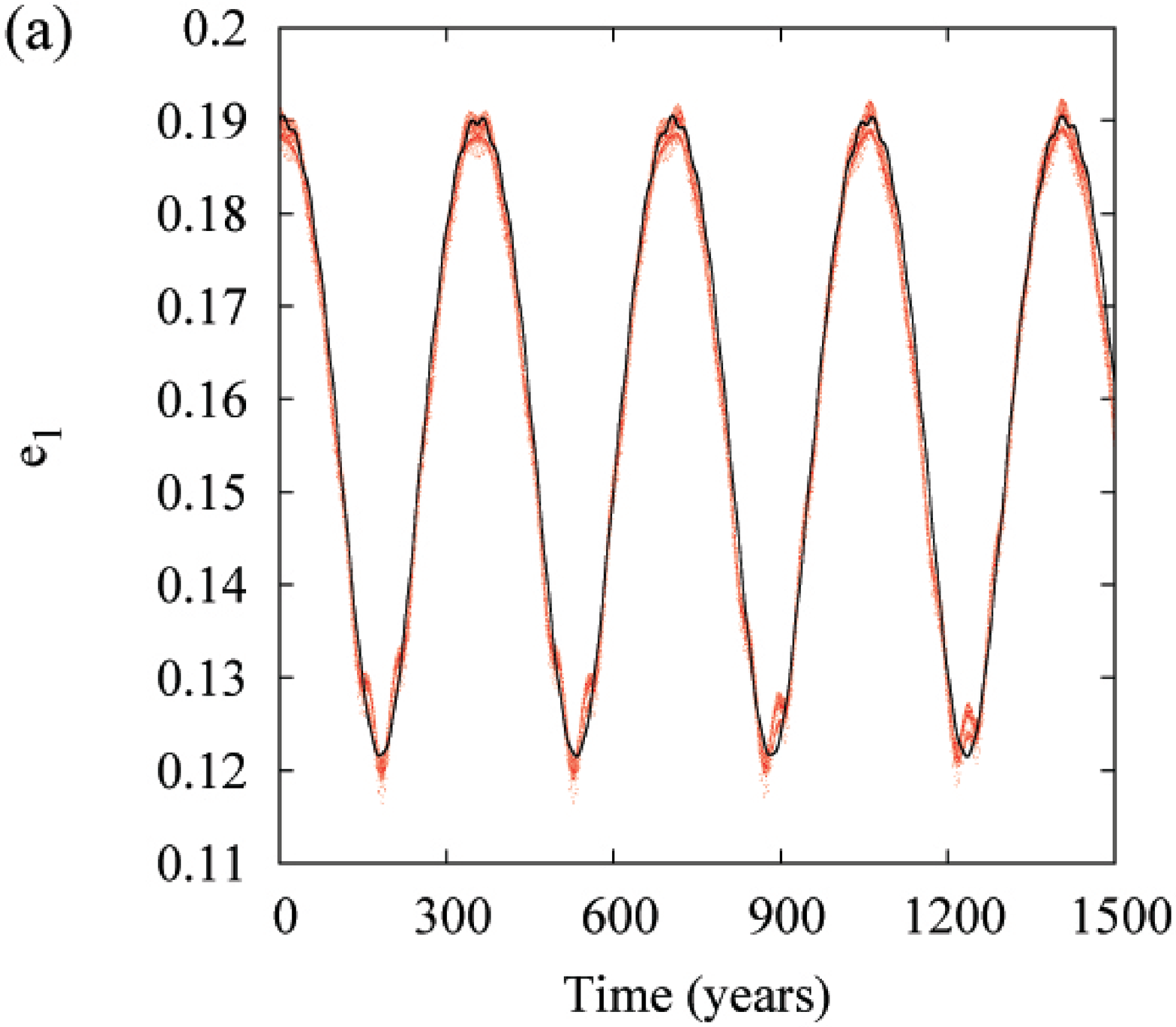}
     \includegraphics[angle=270,scale=0.28]{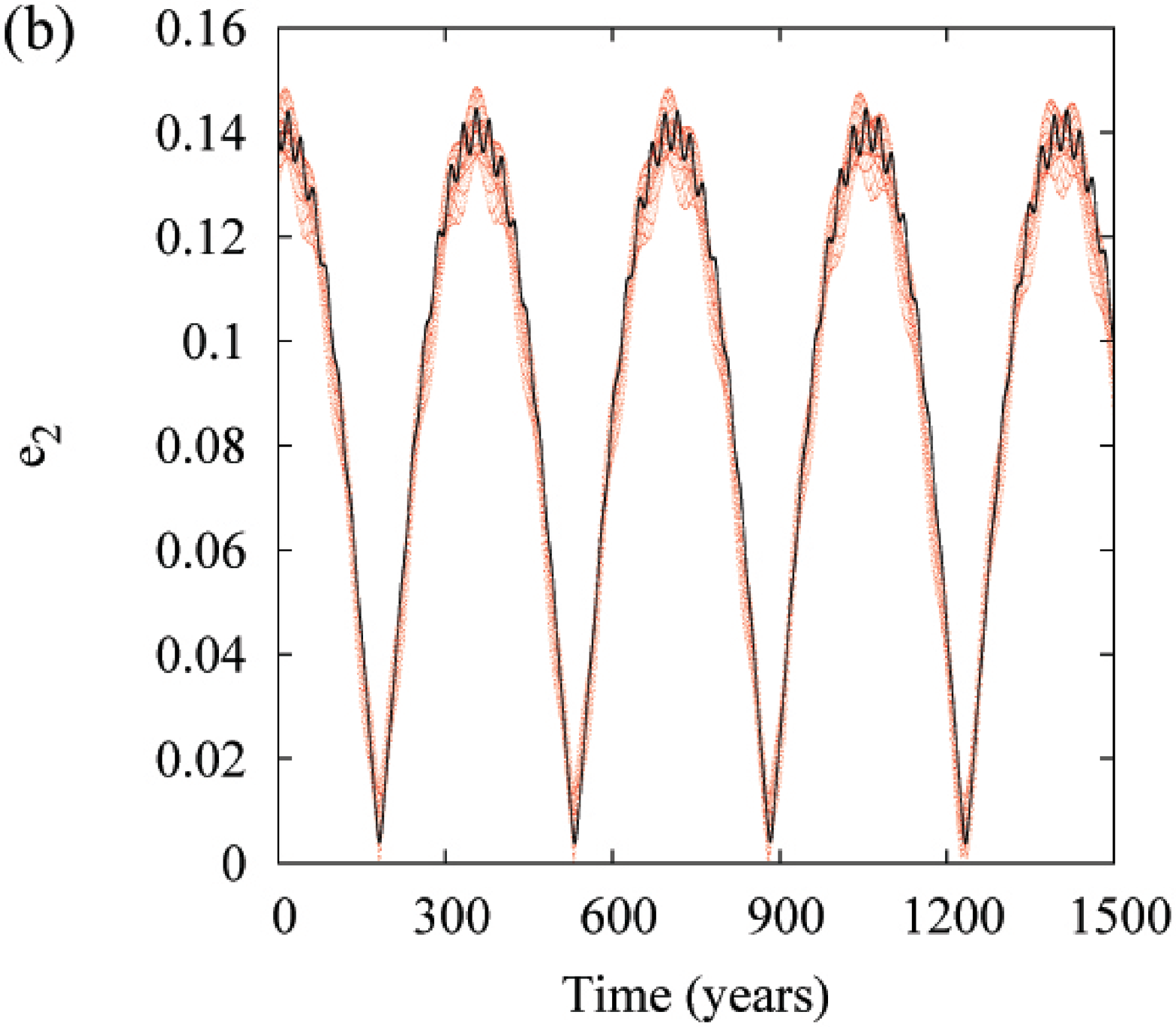}\\
 \caption{Time variation of the eccentricites $e_1$ and $e_2$ 
   for the HD\thinspace73526 planetary system in 2:-1 MMR. Red dots represent 
   the numerical solution while black curves the analytical one. Used initial
   conditions are given by: $M_0 =1.08 \, M_\odot$ ; $M_1 = 2.9 \, M_{Jup}$
   ; $M_2 = 2.5 \, M_{Jup}$ ; $a_1 = 0.66$ ; $a_2  = 1.05$ ; 
   $e_1  = 0.19$ ; $e_2  = 0.14$ ; $\sigma_1  = 94$ ; $\sigma_2  = 94$.}\label{fig_RMMR_e}
\end{figure}

Fig. \ref{fig_RMMR_e} shows the time evolution of eccentricity of both
planets. Dots represent the numerical solution while black curves the
analytical one.\footnote{More details on the used numerical method and
corresponding results
may be found in Gayon \& Bois (2008a).} Both solutions
express the same behavior, with a relative error of $1.2\%$ and $10\%$, on
average, over the eccentricity of the inner orbit and the outer orbit,
respectively (absolute errors being $0.002$ for $e_1$ and $0.005$ for $e_2$ on
average).  
Due to the averaging of the Hamiltonian over short periods, 
numerical dots are scattered on both sides of the analytical solution.

\begin{figure}[!ht]
    \centering
     \includegraphics[scale=0.32]{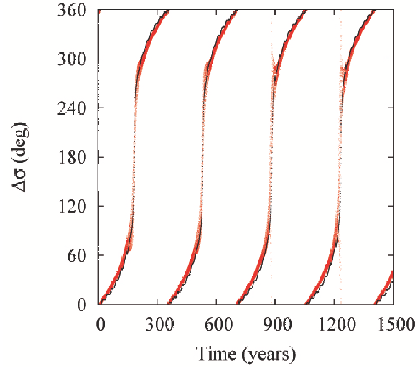}\\
 \caption{Time variation of the $\Delta\sigma$ variable (where
     $\Delta\sigma = \sigma_1 - \sigma_2$)
   for the HD\thinspace73526 planetary system in 2:-1 MMR. Red dots represent 
   the numerical solution while black curves the analytical one. Used initial
   conditions are the same as in 
   Fig. \ref{fig_RMMR_e}.}\label{fig_RMMR_Ds}
\end{figure}

Fig. \ref{fig_RMMR_Ds} shows the time variation of the variable 
$\Delta \sigma$. 
Because the $e_2$ eccentricity periodically reaches the zero value, the
numerical method does not always permit to determine the value of the 
$\Delta \sigma$ angle. This phenomenon is
expressed by the vertical scattering of dots from $0$ to $360$ degrees. As a
consequence, surfaces of section obtained numerically comprise this numerical
bias. The analytical method is therefore more reliable 
when the eccentricity of a planet reaches values close to zero. 
Besides, the analytical approach ensures the properties of stability found
for each planetary system.

%\newpage

\section{Surfaces of section}\label{sec_surfsec} 
Nature of planetary systems can be 
directly deduced from the behavior of the resonance variables.  
As a consequence, from the previous Hamiltonian expansion (\ref{final_Ham}) of 
the three-body problem, we can plot surfaces of section and study the dynamics
of two planets in retrograde MMR.

\begin{figure}[!h]
    \centering
     \includegraphics[angle=270,scale=0.28]{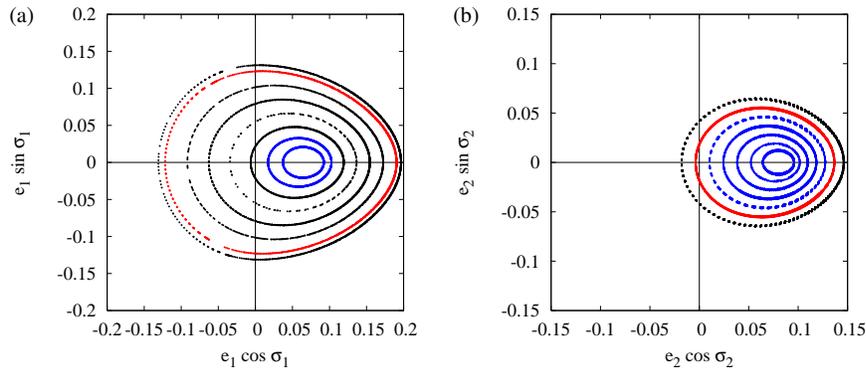}\\
 \caption{Surfaces of section for the inner (a) and the outer (b) planets of
   the HD\thinspace73526 system found close to a 2:-1 MMR. The
   value of $\bar{F}$ is $-0.138352$. Initial conditions (\ref{eq_CI}) are
   represented by the red curve. Black and blue curves respectively 
   correspond to circulation and libration of the $\sigma_i$ variables in the
   plane $\sigma_{j \neq i}=0$.}\label{fig_sect}
\end{figure}

Fig. \ref{fig_sect} shows surfaces of section plotted for the
HD\thinspace73526 planetary system previously defined. 
The energy level corresponding to the initial conditions (\ref{eq_CI}) was
numerically evaluated as $-0.13835250$ (units of AU, solar mass, and
year). With our analytical expansion, we obtain the Hamiltonian value of
$-0.13835197$, which is in good agreement with our numerical results. 
Panel (a) of Fig. \ref{fig_sect} corresponds to the inner planet and is
plotted in the $(e_1\,cos \, \sigma_1\, , \, e_1\,sin \, \sigma_1)$
parameter space. The section plane for the inner planet is chosen such as
$\sigma_2=0$. Similarly, panel (b) is plotted for the outer planet in the 
$(e_2\,cos \, \sigma_2\, , \, e_2\,sin \, \sigma_2)$ parameter space and using
the plane $\sigma_1=0$. The initial conditions corresponding to  
(\ref{eq_CI}) are plotted in red. All the solutions found for
$\bar{F}=-0.138352$ are quasi-periodic.

In panels (a) and (b) of Fig. \ref{fig_sect}, we represent the 
{\it circulation} state of the $\sigma_1$ and $\sigma_2$ variables in black 
and red colors. The {\it libration} state about $0^\circ$ is plotted in blue; 
such initial conditions are located {\it inside} the $2:1$ retrograde MMR. 
Nevertheless, for all the sets of initial conditions required for plotting 
Fig. \ref{fig_sect}, the time variation of $\sigma_1$ and $\sigma_2$ expresses 
a {\it circulation} (in time) of both variables (not shown here). This time 
circulation seems contrary to the possible behavior in libration of 
$\sigma_1$ and $\sigma_2$ shown on our surfaces of section. Such initial 
condition sets inside the $2:1$ retrograde MMR and their behaviors 
could therefore point out a particular characteristic of retrograde MMR. 
Such a characteristic would deserve a specific study.
 
\vspace{1cm}

\newpage
\section{Conclusion}
In the present paper, we investigated the three-body problem in the particular
case of retrograde resonances. Our study is derived from the Hamiltonian 
approach of Beaug\'e \& Michtchenko (2003), which was expanded for eccentric, 
coplanar and prograde orbits. To apply this method to the retrograde MMR case,
we expanded again the disturbing function, when considering a prograde motion
of the inner planet and a retrograde motion of the outer planet. Hence, we
defined a new set of canonical variables, which allow us to express correctly 
the angles of resonance in the case of counter-revolving configurations.

Although the exploration of the {\it N}-body problem is accessible from 
numerical methods, the acquiring of an analytical ``rail'' notably contributes
to a deeper understanding of the numerical investigation. As shown in Section 
\ref{compare}, the analytical method also permits a better determination of 
the resonance variables. Moreover, from the behavior of the resonance angles 
displayed in surfaces of section, we can directly infer the local behavior of 
a 3-body system, that is to say its stability or its chaoticity.  

Until now, no planetary system has been truly detected in counter-revolving 
configuration. However, given the efficiency for stability of retrograde MMR 
(see Gayon \& Bois 2008b), such a detection might occur. The work presented in 
this paper is firstly based on a dynamical study of theoretical two-planet 
systems. Nevertheless, a dynamical system composed of two satellites orbiting
a planet is an equivalent problem. Since some satellites of Saturn are found in
counter-revolving motions, our expansion of the three-body problem solved in 
the case of retrograde motions could then be applied for satellites of the
Solar System.   

\begin{acknowledgements}
We thank the anonymous referees for their  useful comments and suggestions
that  helped to improve the paper. 
\end{acknowledgements}

\begin{appendix}
\section{Appendix}
This appendix shows the analytical expansion when the body 
moving on a retrograde orbit is the inner planet. The new set of canonical 
variables for the resonant averaged Hamiltonian is written as follows 
(equivalent to \ref{eq_average})~:

\begin{equation}
   \begin{aligned}
   \lambda_1 & 
   & J_1 =& -L_1 + s (I_1 + I_2)\\
   \lambda_2 & 
   & J_2 =&  \hspace{4mm}L_2 - (1+s) (I_1 + I_2)\\
   \sigma_1  & = (1+s) \lambda_2 - s \lambda_1 - \tilde{\omega}_1 \hspace{1cm}
   & I_1 =& -L_1\, (1 - \sqrt{1-e_1^2})  \\
   \sigma_2 &= (1+s) \lambda_2 - s \lambda_1 - \tilde{\omega}_2  
   & I_2 =& \hspace{4mm}L_2\, (1-\sqrt{1-e_2^2}) 
   \end{aligned}
\end{equation}
with $\lambda_1=-M_1+\tilde{\omega}_1$, $\lambda_2=-M_2-\tilde{\omega}_2$, 
$\tilde{\omega}_1=\Omega_1-\omega_1$ and $\tilde{\omega}_2=\Omega_2+\omega_2$.

Although the set of canonical variables changes, the
expression of the resonant averaged Hamiltonian remains the same~:
\begin{equation}
\bar{F} = -\sum_{i=1}^{2} \frac{\mu_i^2 \beta_i^3}{2 L_i^2} 
      - \frac{G m_1 m_2}{a_2} \sum_{i,j,k,m,l} \bar{R}_{i,j,k,m,l} \, \alpha_i \,
      e_1^j \, e_2^k \, \cos((m-l)\sigma_1 + (l-n) \sigma_2)
\end{equation}
\end{appendix}

{}

\end{document}